\documentclass[copyright,creativecommons,noderivs]{eptcs}
\providecommand{\event}{SYNT 2012}
\usepackage[latin1]{inputenc}
\usepackage{amsmath}
\usepackage{amssymb}
\usepackage{underscore}
\usepackage{amsthm}
\usepackage{tikz}
\usepackage{longtable}
\usetikzlibrary{automata,patterns,decorations,decorations.pathmorphing}

\newcommand{\NN}{\mathbb{N}}

\newcommand{\newterm}{\emph}
\newcommand{\Succ}{\mathrm{succ}}
\newcommand{\LTLG}{\mathsf{G}}
\newcommand{\LTLF}{\mathsf{F}}
\newcommand{\LTLX}{\mathsf{X}}

\title{Sparse Positional Strategies for Safety Games\thanks{This work was partially supported by the DFG as
part of the Transregional Collaborative Research Center ``Automatic Verification and Analysis of Complex
Systems'' (SFB/TR 14 AVACS).}}
\author{R\"udiger Ehlers $\quad \quad$ Daniela Moldovan
\institute{
Reactive Systems Group\\Saarland University\\66123 Saarbr\"ucken, Germany}}
\def\titlerunning{Sparse Positional Strategies for Safety Games

}
\def\authorrunning{R. Ehlers \& D. Moldovan}
\begin{document}
\maketitle

\begin{abstract}
We consider the problem of obtaining sparse positional strategies for safety games. Such games are a commonly used model in many formal methods, as they make the interaction of a system with its environment explicit. Example applications are the synthesis of finite-state systems from specifications in temporal logic and alternating-time temporal logic (ATL) model checking. Often, a winning strategy for one of the players is used as a certificate or as an artefact for further processing in the application. \emph{Small} such certificates, i.e., strategies that can be written down very compactly, are typically preferred. For safety games, we only need to consider positional strategies. These map game positions of a player onto a move that is to be taken by the player whenever the play enters that position. For representing positional strategies compactly, a common goal is to minimize the number of positions for which a winning player's move needs to be defined such that the game is still won by the same player, without visiting a position with an undefined next move. We call winning strategies in which the next move is defined for few of the player's positions \emph{sparse}.
From a sparse winning positional strategy for the safety player in a synthesis game, we can compute a small implementation satisfying the specification used for building the game, and for ATL model checking, sparse strategies are easier to comprehend and thus help in analysing the cause of a model checking result. 

Unfortunately, even roughly approximating the density of the sparsest strategy for a safety game has been shown to be NP-hard. Thus, to obtain sparse strategies in practice, one either has to apply some heuristics, or use some exhaustive search technique, like ILP (integer linear programming) solving. In this paper, we perform a comparative study of currently available methods to obtain sparse winning strategies for the safety player in safety games. Approaches considered include the techniques from common knowledge, such as using ILP or SAT (satisfiability) solving, and a novel technique based on iterative linear programming. The restriction to safety games is not only motivated by the fact that they are the simplest game model for continuous interaction between a system and its environment, and thus an evaluation of strategy extraction methods should  start here, but also by the fact that they are sufficient for many applications, such as synthesis.  
The results of this paper shed light onto which directions of research in this area are the promising ones, and if current techniques are already scalable enough for practical use. 

\end{abstract}

\section{Introduction}
\label{sec:introduction}

Games with $\omega$-regular winning conditions have been proven to be valuable tools for the construction and analysis of complex systems and are suitable computation models for logics such as the monadic second-order logic of one or two successors \cite{DBLP:conf/dagstuhl/2001automata,DBLP:journals/igpl/Stirling99,DBLP:journals/jacm/AlurHK02,Thomas1994}. By reducing a decision problem to determining the winning player in a game, the algorithmic aspect of solving the problem can easily be separated from the details of the application under concern. Winning strategies for one of the players in a game can be used as \newterm{certificates} for the answer to the original problem or serve as \newterm{artefacts} to be used in other steps of the application.

For example, when synthesizing finite state systems \cite{Vardi1996,DBLP:journals/bsl/KupfermanV99} from temporal logic specifications, the winning strategy for the system player in the corresponding game is an artefact that represents a system satisfying the specification, and is used for building circuits that implement the specification. In alternating-time temporal logic (ATL) \cite{DBLP:journals/jacm/AlurHK02}, the question is imposed whether agents in a certain setting can ensure certain global properties of a system to hold. A winning strategy for one of the players in the corresponding model checking game represents a certificate for the fact that the agents can achieve their goal or that there exists a counter-strategy for the remaining agents to prevent this. The certificate can then be used for human inspection on why or why not the agents can achieve their goal. In $\mu$-calculus model checking, a strategy for the induced model checking game explains why or why not a given system satisfies some property. Again, a winning strategy serves as a certificate that is useful for further analysis of the setting.

In all these cases, certificates and artefacts that have a smaller representation are normally preferred. Such solutions are easier to comprehend and have (computational) advantages if used in successive steps (like building circuits from strategies in a synthesis game) or for analysing why a certain property holds or not. While for automata over $\omega$-regular words, which can be seen as  one-player games, there exist some results on obtaining compactly representable one-player strategies for Büchi \cite{DBLP:conf/concur/KupfermanS06} and generalised Büchi \cite{DBLP:conf/dac/ClarkeGMZ95} acceptance conditions, little research has been performed on obtaining compactly representable strategies in two-player games, even though it has been noticed that these are desperately needed in practice \cite{DBLP:journals/entcs/BloemGJPPW07}.

In this paper, we consider the problem of obtaining \newterm{sparse} \newterm{positional} strategies in safety games. Whenever a player follows a positional strategy, then the choice of action to perform in one of its positions only depends on the position the game is currently in. While positional strategies are too restricted to allow representing winning strategies in very expressive game types such as Muller or Streett games in general, for more simple game types such as parity or safety games, it is assured that whenever for one of the players, a winning strategy exists, then there also exists a winning positional strategy for that player. Positional strategies are suitable for giving insight on why a modal $\mu$-calculus formula is valid in some model or provide information about why a specification is unrealisable in synthesis, as the obligations are encoded into the game graph. Technically, positional strategies are represented as functions from the positions of a player to the next move of the player. Thus, at a first glance, all strategies have the same size. However, if some position is never reachable along a play, then the player's move at that position does not matter, and we can leave the move for this position \newterm{undefined}. Positional strategies with many undefined moves can be represented more compactly, have the advantages outlined above, and are what we aim at computing in this paper. The number of game positions for which a next move is defined in a positional strategy is called its \newterm{density}, and strategies with a low density are called \newterm{sparse} in the following.

For applications such as synthesis, positional strategies are not necessarily the best model: a Mealy or Moore machine that implements a specification can have far less states than the density of the sparsest winning positional strategy for the system player in a corresponding synthesis game. Nevertheless, even in synthesis, positional strategies are useful. For example, one of the more recent synthesis approaches, namely \emph{Bounded Synthesis} \cite{DBLP:conf/atva/ScheweF07a}, can easily be altered such that there exists a positional strategy that represents the smallest possible implementation. Furthermore, stronger non-approximability results are known for non-positional strategies: it was shown that the number of states of the smallest Mealy or Moore machine that implements a winning strategy in a safety game is NP-hard to approximate within any polynomial function \cite{DBLP:conf/lata/Ehlers10}, while for positional strategies, non-approximability of the density of a sparsest strategy is known only within any constant \cite{atr080}. 

While safety games are the main computational model that we aim to tackle, the techniques we compare in this paper are also useful for more expressive game models such as parity games. Extracting winning strategies in parity games can be done by computing a strategy that follows some attractor sets computed during the game solving process \cite{DBLP:conf/focs/EmersonJ91}. If we leave the concrete choice of a successor position in such a game open whenever there is more than one possibility to follow the attractor, we obtain a \emph{non-deterministic} strategy that leaves some room for density improvement. 
Any strategy that is a special case of this non-deterministic strategy is a valid winning strategy, just like every strategy that does not leave the set of \newterm{winning} positions in a safety game is a valid winning strategy. Thus, the techniques discussed here can also be applied to the parity game case, with the drawback that the sparsest winning strategy in a parity game might not be a special case of the non-deterministic strategy computed from the attractor sets observed during the game solving process, and thus may be missed. Nevertheless, as there is, to the best of our knowledge, no work on sparse strategies in parity games yet, using an approach to obtain sparse winning strategies in safety games is still the best technique available so far. 

We compare a variety of techniques for obtaining sparse winning strategies in this paper. Apart from a fully randomized heuristic, which will serve as a comparison basis, we use a smarter randomized heuristic that finds locally optimal strategies and consider the usage of SAT and ILP solvers to obtain a sparsest strategy. A novel technique, based on the repeated application of a linear programming solver to obtain hints on which game position to add to the strategy domain next provides a trade-off between the density of the strategy and the computation time needed. For comparison, we also consider a recent algorithm by Neider \cite{DBLP:conf/atva/Neider11}, which uses computational learning to obtain small non-positional strategies. As there is no standard benchmark set available for safety games, we take games from the Bounded Synthesis domain. 

We start the following presentation by defining safety games. As we compare the techniques to obtain sparse positional winning strategies against the computational learning approach, which  produces non-positional strategies, we use an action-based definition of safety games, which ensures that the strategy types stand on a common ground. In Section \ref{sec:approaches}, we describe the techniques considered to find sparse strategies. Then, in Section \ref{sec:benchmarks}, we briefly describe the benchmarks used. Preceded by a short description of the experimental setup (including the tools used), we then state the experimental results in Section \ref{sec:experiments}. We close with a discussion of the results and indicate open problems. 

Due to space restrictions, we do not describe how the computational learning-based strategy finding approach \cite{DBLP:conf/atva/Neider11} works and how to produce games from specifications in the Bounded Synthesis \cite{DBLP:conf/atva/ScheweF07a} process. Rather, we assume familiarity with the subjects in the corresponding sections \ref{sec:neiderLearning} and \ref{sec:BoundedSynthesis}, and only explain the connection to this work. 

\section{Preliminaries}
\label{sec:definitions}

\subsection{Safety Games}
A safety game is defined as a tuple $\mathcal{G} = (V^0,V^1,\Sigma^0,\Sigma^1,E^0,E^1,v^\mathit{init})$. In the game, we have two competing players, namely player $0$ and player $1$. Player $0$ has the (finite) set of positions $V^0$, the (finite) set of actions $\Sigma^0$, and the partial edge function $E^0 : V^0 \times \Sigma^0 \rightharpoonup (V^0 \uplus V^1)$. Player $1$ in turn has her set of positions $V^1$, her set of actions $\Sigma^1$ and her edge function $E^1 : V^1 \times \Sigma^1 \rightharpoonup (V^0 \uplus V^1)$. The game also has a designated initial position $v^\mathit{init} \in (V^0 \uplus V^1)$. For simplicity, we define $V = V^0 \uplus V^1$, $\Sigma = \Sigma^0 \uplus \Sigma^1$, and $E : V \times \Sigma \rightharpoonup V$ with $E(v,x) = E^0(v,x)$ if $E^0(v,x)$ is defined and $E(v,x) = E^1(v,x)$ otherwise as shortcuts to be used in the following. If for some position $v \in V$ and action $x \in \Sigma$, we have $E(v,x) = v'$ for some position $v'$, then we call $v'$ a \newterm{successor} of $v$. The set of successors of a position $v$ is also denoted by $\Succ(v)$.

In a \emph{play} of the game, the players move a pebble along the positions in the game. Starting from the initial position, whenever the pebble is in a position $v \in V^0$, then player $0$ chooses an action $x \in \Sigma^0$ and moves the pebble to position $E^0(v,x)$. The case of the pebble being in a position $v \in V^1$ is analogous for player $1$. By concatenating the actions taken by the two players along the play, we obtain a \emph{decision sequence} in the game. 

Given a set $X$, we denote the set of finite sequences of $X$ by $X^*$, and the set of infinite sequences of $X$ by $X^\omega$.
A sequence $\pi = \pi_0 \pi_1 \pi_2 \ldots \in V^\omega \cup V^*$ is then a play with a corresponding decision sequence $\rho = \rho_0 \rho_1 \rho_2 \ldots \in \Sigma^\omega$ if $\pi_0 = v^\mathit{init}$ and for all $i \in \NN$, if $\pi_i \in V^0$, then $\rho_i \in \Sigma^0$ and $\pi_{i+1} = E^0(\pi_i,\rho_i)$ (or $i = |\pi|-1$ if $E^0(\pi_i,\rho_i)$ is undefined), and if $\pi_i \in V^1$, then $\rho_i \in \Sigma^1$ and $\pi_{i+1} = E^1(\pi_i,\rho_i)$ (or $i = |\pi|-1$ if $E^1(\pi_i,\rho_i)$ is undefined). Note that for every decision sequence, there is precisely one play to which it corresponds.
Plays in a game are either winning for player $0$ or player $1$. Finite plays $\pi = \pi_0 \ldots \pi_n$ for which we have $\pi_n \in V^0$ are winning for player $1$, whereas for $\pi_n \in V^1$, the play is winning for player $0$. Infinite plays are won by player $0$. 

\subsection{Strategies}

When playing the game, a player may follow a predefined \emph{strategy}. Formally, a strategy for player $p \in \{0,1\}$ is simply a function $f : \Sigma^* \rightarrow \Sigma^p$. A decision sequence $\rho$ is said to correspond to $f$ if for the play $\pi$ that $\rho$ corresponds to and all $i \in \NN$, if $\pi_i \in V^p$, then $\rho_i = f(\pi_0 \ldots \pi_{i})$. If all decision sequences that correspond to a given strategy of player $p$ induce only plays that are winning for player $p$, then we call the strategy $f$ \newterm{winning}.

In safety games, it is assured that one of the two players has a winning strategy (see, e.g.,  \cite{DBLP:conf/dagstuhl/2001automata}). If player $p$ has a strategy to win the game, then we say that player $p$ \emph{wins the game}. We can restrict our attention to a special kind of strategies, namely \emph{positional strategies}. We call a strategy $f : \Sigma^* \rightarrow \Sigma^p$ positional if
for all pairs of prefix decision sequences $\rho = \rho_0 \ldots \rho_n$ and $\rho' = \rho'_0 \ldots \rho'_m$, 
if $E( \ldots E(v^\mathit{init},\rho_0), \ldots, \rho_n) = E( \ldots E(v^\mathit{init},\rho'_0), \ldots, \rho'_m)$, then $f(\rho) = f(\rho')$. In other words, at any position in a play, the next decision of a player that follows a positional strategy only depends on the position the play is in at that time. As a consequence, such a positional strategy can also be described by a function $f : V^p \rightarrow \Sigma^p$ that maps every position of player $p$ in the game to an action to be chosen by the player whenever the position is visited. The restriction to positional strategies is motivated by the fact that in safety games, whenever there exists a winning strategy for one of the players, then there also exists a positional strategy for the player. The standard algorithm to solve safety games (i.e., determining the winner of the game) described in the next sub-section also produces positional strategies as certificates/artefacts.

For comparing different strategies and in particular finding sparse strategies, we need to define a density measure for positional strategies. Recall that the motivation of focusing on sparse strategies is that they are better comprehensible certificates and have computational advantages when used as artefacts for further processing. For positional strategies, we only need to consider choices from positions in $V^p$ that are reachable along some path that corresponds to the strategy. If for a positional strategy $f : V^p \rightarrow \Sigma^p$, there is some position $v \in V^p$ that can never be reached along a path that corresponds to a decision sequence that in turn corresponds to the strategy, then for the positional strategy $f$, $f(v)$ can be arbitrary without changing the behaviour of the strategy. We thus define the density of a positional strategy for player $p$ to be the number of positions of player $p$ that can be visited along some play that corresponds to this strategy. 

More formally, we could also define $f$ as a \emph{partial} function from $V^p$ to $\Sigma^p$ and define the strategy density to be the size of the domain of $f$. In this case, whenever the pebble is in a position $v^p \in V^p$ for which $f(v^p)$ is undefined, we assume that player $p$ declares that she loses the play. If the strategy is still winning under this modified definition of who wins a play, then the fact that $f$ is only a partial function apparently does not to matter, and $f$ can be considered to be a valid positional strategy.

\subsection{Solving Safety Games}

For discussing the problem of obtaining sparse positional strategies in safety games, it is reasonable to separate the complexity of the process of solving the game (which is doable in polynomial time) from the actual optimization problem of minimizing the strategy (which is NP-hard). Solving the game means to identify the set of \newterm{winning} positions in the game, i.e., those for which if any of these positions is an initial one, the safety player (player $0$) wins the game. Solving a safety game is relatively simple: it can be shown that the set of winning positions is precisely the largest set of positions that (1) does not contain a position of player $0$ that has no successors, (2) for which for every position of player $0$, one of its successors is in the set, and (3) for every position of player $1$, all of its successors are in the set. This largest set can be computed by starting with all positions, and successively removing any position that does not satisfy (1), (2), or (3). Once no more positions can be removed, the game solving process is complete. 

Let $W$ be the set of winning positions and $v^\mathit{init} \in W$. Any positional strategy $f : V^0 \rightarrow \Sigma^0$ for which for all $v \in V^0 \cap W$, we have that $E^0(v,f(v)) \in W$, is a winning one, as it ensures that the set of winning positions is not left, by condition (2) above, player $1$ cannot initiate leaving $W$ along a play, and no dead end for player $0$ is part of $W$. At the same time, any positional strategy that allows leaving $W$ at some point in a play is not winning. This motivates the description of a \newterm{most permissive winning strategy} for player $0$ in the game: we define $f' : V^0 \rightarrow 2^{\Sigma^0}$ with $f'(v) = \{ x \in \Sigma^0 \mid E^0(v,x) \in W\}$ for every $v \in V^0$, as every concrete winning positional strategy must be a specialization of $f'$, i.e., have $f(v) \in f'(v)$ for every position $v \in V^0 \cap W$ that is reachable along some play that corresponds to $f$. 
For a procedure to find sparse positional strategies in a safety game, we can thus use $f'$ as a basis for finding a sparse specialization.

\section{Approaches for Obtaining Sparse Winning Strategies}
\label{sec:approaches}

In the experimental evaluation to follow, we compare five techniques to obtain sparse winning strategies in games. In this section, we explain them and state the properties of the approaches. We are particularly interested in \newterm{sparsest} strategies in safety games, i.e., winning positional strategies with the lowest possible density. 

\subsection{Randomized Strategy Extraction}
Probably the most simple way to obtain a concrete winning positional strategy from a most permissive strategy is to simply pick arbitrarily one allowed action for every winning position of player $0$, and then to remove all positions that became unreachable from the strategy domain. Here, we perform a random pick, based on a uniform distribution over the available actions. 

\subsection{Smarter Randomized Strategy Extraction}
Given a game $\mathcal{G}=(V^0,V^1,\Sigma^0,\Sigma^1,E^0,E^1,v^\mathit{init},\mathcal{F})$ with the set of winning position $W$ for player $0$ (and $v^\mathit{init} \in W$), another way to describe the problem of obtaining sparse winning positional strategies is to search for an as-large-as-possible set of positions $Z \subset W$ for which the concrete positional strategy function should be undefined. Any strategy that respects $Z$ will then have the same density (as otherwise, there is some position that we can add to $Z$ and thus $Z$ is not as large as possible). 

As finding the density of sparsest positional strategies is NP-hard to approximate within any constant \cite{atr080}, finding an approximately largest set $Z$ is also NP-hard. However, we might settle for \emph{local optima} of $Z$, i.e., declare ourselves to be satisfied to obtain a set $Z$ such that there is no position of player $0$ in the game that can be added to $Z$ such that there is still a winning positional strategy that respects $Z$ (i.e., has $f(v)$ undefined for every $v \in Z$). Such a set can be obtained in  time polynomial in the size of the game (i.e., in $|V^0| \cdot |\Sigma^0| +  |V^1| \cdot |\Sigma^1|$).

In particular, we can do so as follows: we first create a random permutation of $V^0$, and then for every position in the list, examine if the safety game is still winning for player $0$ if we remove all outgoing edges of that position. Whenever this is the case, we add the position to $Z$, and continue. Whenever the safety game becomes losing for player $0$ with this change, we undo it and try the next position in the list. Once every position in the list has been tried, we obtained a locally optimal set $Z$ (whose local maximality easily be proven by deriving a contradiction from assuming the converse).

Since we randomize the permutation, for every game, there is a non-zero probability of obtaining a sparsest strategy. However, it is not hard to define a series of games $\mathcal{G}_1, \mathcal{G}_2, \ldots$ for which the sizes of the games $\mathcal{G}_i$ grow linearly in $i$, but for which the probability to obtain a sparsest strategy using the algorithm above is at most $\frac{1}{2^i}$ for every game $\mathcal{G}_i$.

\subsection{Integer Linear Programming}
\label{sec:ilp}
Given a game $\mathcal{G}$, we can formulate the problem of obtaining a sparse positional winning strategy for player $0$ as an integer linear programming (ILP) problem, in which we use one variable per position in the game. Whenever we obtain a solution to the ILP problem, a variable value of $1$ is supposed to mean that the position can be reached from the initial position along some path that corresponds to the computed strategy, whereas a value of $0$ means the opposite. By optimizing the sum of the variables that correspond to the vertices of $V^0$, we can obtain a sparsest strategy.

Formally, an ILP problem is a three-tuple $\langle X, F, C \rangle$, for which $X$ is a set of variables, $F$ is a linear function over $X$ that is to be minimized, and $C$ is a set of linear constraints over the allowed values of $X$. Given a game $\mathcal{G} = (V^0,V^1,\Sigma^0,\Sigma^1,E^0,E^1,v^\mathit{init},\mathcal{F})$ and a most permissive strategy $f : V^0 \rightarrow 2^{\Sigma^0}$, we can encode the problem of obtaining a sparsest positional winning strategy for player $0$ that is a specialization of $f$ into an ILP problem $\langle X, F, C \rangle$ by setting $X = V$, $F = \sum_{v \in V^0} v$, and:
\begin{align*}
 C & = \bigcup_{v \in V} \{ v \geq 0, v \leq 1 \} \cup \{ v^\mathit{init} \geq 1 \} \\
  & \cup \{ - v + \sum_{x \in f(v)} E^0(v,x) \geq 0 \mid v \in V^0 \} 
  \cup \{ -v + v' \geq 0 \mid v \in V^1, v' \in \Succ(v) \}.
\end{align*}
There are four types of constraints in this ILP formulation: first of all, all variable values are fixed between 0 and 1. Then, the variable corresponding to the initial position in the game is forced to be $1$. For every position of player $0$ whose variable value is $> 0$, the third kind of constraint ensures that the variable for \emph{some} successor position that is reachable via some action allowed by $f$ has to be set to 1. Finally, for positions of player $1$ whose variable has a value of $> 0$, the variables for all successors positions have to be $1$.
For actually obtaining a positional strategy from a variable assignment $a : X \rightarrow \{0,1\}$, for every position, we pick an action that leads to a successor in $\{ v \in V | a(v) = 1\}$.

\subsection{SAT-based Strategy Extraction}

The ILP formulation of the sparsest positional strategy problem has the property that when regarding the variables as Boolean by interpreting $0$ as $\mathbf{false}$ and $1$ as $\mathbf{true}$, all of the constraints can be represented as a disjunction of Boolean literals. For example, a constraint $-v_1 + v_2 + v_3 \geq 0$ can be written as $\neg v_1 \vee v_2 \vee v_3$ in the Boolean domain. By rewriting the ILP instance in this way, SAT (satisfiability) solvers can be applied. A SAT-based approach to strategy finding has already been pursued in \cite{Ellonen2010}.

Most currently available SAT solvers however cannot take into account optimization objectives when computing a solution. For using such a solver then, we could encode some cardinality constraint on the amount of variables for player $0$'s positions that might be set to $\mathbf{true}$ at most, and perform a binary search on the best possible strategy density. For this paper, we use the SAT solver \textsc{OPTSAT v.1.1} \cite{DBLP:conf/jelia/GiunchigliaM06} that has this functionality already built in.

\subsection{Repetitive Linear Programming}

\looseness-1 The integer linear programming approach from Section \ref{sec:ilp} is exact and guaranteed to find a sparsest strategy. As the problem of obtaining sparsest winning positional strategies is NP-hard, we cannot expect ILP solvers to work fast on ILP instances that encode this problem in general. To counter this fact, we propose an alternative approach here, which implements a heuristic based on linear programming over the real numbers (LP). In contrast to ILP solving, LP solving can be performed in polynomial time.

Consider the constraint system built in the ILP approach of Section \ref{sec:ilp}, but this time over the real numbers. After applying a linear programming solver to the system, we obtain a variable valuation $\vec v$, which is, w.l.o.g., of the form $(v_1,v_2, \ldots, v_m)$ for $m = |V|$. Some values here might be $0$, some might be $1$ and in many cases, some values are in between. Thus, the values might not represent an actual solution to the sparse strategy problem. We can however fix the vector in an iterative fashion. Suppose that we start the linear programming solver on the problem again, but this time fix all variables that were $0$ in $\vec v$ after the previous solver run to $0$, fix all variables that had a value of $1$ in $\vec v$ to $1$, and additionally fix one variable whose value was equal to $\max \{ v_i \mid 1 \leq i \leq m, v_i \neq 1\}$ to $1$. The linear programming solver will compute a new solution, but possibly with a worse value of the objective function. However, the number of variables that are not $0$ or $1$ will have decreased by at least $1$. If we iterate the process until all variables have values of either $0$ or $1$, we have a blueprint for a sparse, but not necessarily sparsest strategy. However, the complexity of this approach is only polynomial, and we use the LP solver to guide our search for a sparse winning strategy.

\subsection{Computational Learning of Sparse Strategies}
\label{sec:neiderLearning}
\begin{figure}
\centering \begin{tikzpicture}
  \path[use as bounding box] (-0.9,-0.76) rectangle (13.5,1.23); 
  \node[accepting,state,minimum size=0.7cm,shape=circle] (a) at (0,0) {$q_0$};
  \node[accepting,state,minimum size=0.7cm,shape=circle] (b) at (2,0) {$q_1$};
  \node[accepting,state,minimum size=0.7cm,shape=circle] (c) at (4,0) {$q_2$};
  \node[accepting,state,minimum size=0.7cm,shape=circle] (d) at (6,0) {$q_3$};
  \draw (-0.9,0) edge[semithick,->] (a);
  \draw (a) edge[semithick,->] node[above] {$u,v$} (b);
  \draw (b) edge[semithick,->] node[above] {$y,z$} (c);
  \draw (c) edge[semithick,->] node[above] {$u$} (d);
  \draw[semithick,->] (d) .. controls +(0.2,1.2) and +(-0.2,1.2) .. node[below] {$x$} (a);
  \draw[semithick,->] (c) edge[bend left=25] node[below] {$v$} (b);
  \draw [semithick, ->, decorate, decoration={snake,amplitude=.5mm,segment length=2mm}] (6.8,0) -- (8.6,0);
  \node[state,minimum size=0.7cm,shape=circle] (sa) at (10,0) {$s_0$};
  \node[state,minimum size=0.7cm,shape=circle] (sb) at (12,0) {$s_1$};
  \draw (9.1,0) edge[semithick,->] (sa);
  \draw[semithick, ->] (sa) edge[bend left=15] node[above] {\begin{tabular}{c} $u/z$ \\ $v/z$ \end{tabular}} (sb);
  \draw[semithick, ->] (sb) edge[bend left=15] node[below] {\begin{tabular}{c} $u/x$ \end{tabular}} (sa);
  \draw (sb) edge[->,loop right] node[right] {$v/z$} (sb);
  
  \end{tikzpicture}
\caption{Translating a DFA that represents winning decision sequences in a safety game with $\Sigma^0 = \{x,y,z\}$, $\Sigma^1 = \{u,v\}$, and $v^\mathit{init} \in V^1$ into a Mealy machine. Note that as from $q_1$ in the DFA, there are multiple possible next actions, for the Mealy machine, we just picked any of them (i.e., $z$). Edges in the Mealy machine denote both inputs an outputs. For example, the transition from $s_1$ to $s_0$ is taken when $u$ is read in state $s_1$, and when taking the transition, $x$ is put out.}
\label{fig:makingMealyMachine}
\end{figure}
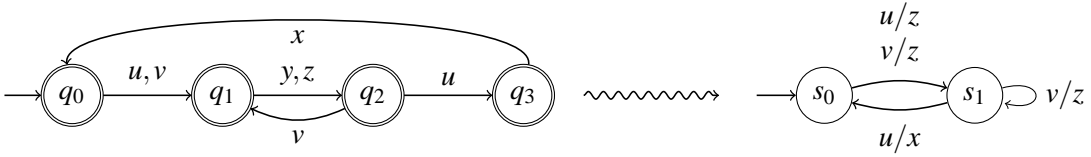

Recently, the problem of obtaining compactly representable winning strategies in safety games has been tackled from a computational learning perspective by Neider \cite{DBLP:conf/atva/Neider11}. In computational learning of a regular language over finite words, the task is to obtain a deterministic finite automaton (DFA) representation of such a language using only equality and containment checks. The idea in applying this idea to strategy extraction is that we use the prefixes of the winning decision sequences for player $0$ in a game as a language to be learned, but we can actually stop the learning process after a subset of this language has been learned that is closed under appending allowed actions of player $1$  (i.e., those actions that are available to player $1$ at a respective point in the game). The left part of Figure \ref{fig:makingMealyMachine} depicts an example automaton for such a language.

Note that the automaton is also concerned with actions of player $1$, and when taking its number of states as size measure, it can easily be larger than the density of a positional strategy. However, at the same time, a strategy automaton can also be smaller, as it allows to merge states with the same suffix language. 
Also, a strategy DFA might offer more than one possible action to player $0$ at any point in the play, and there is no guarantee that there actually exists a \emph{positional} strategy in the game that the DFA represents (or overapproximates). As a consequence, the density of the sparsest positional strategy and the size of the smallest automaton-based representation of a strategy are incomparable.

For games that represent some synthesis problem and have strict alternation between the two players in the game, positional strategies are not necessarily the model of choice. Typically, when the safety player is winning such a game, it is desired to build a Mealy or Moore machine from a winning strategy that then represents a reactive system that satisfies the specification that the game is built from. Such a Mealy or Moore machine takes the actions of the other player as input and produces player 0's actions as output. Any trace that the machine may produce must then be a winning decision sequence in the original game. A Mealy or Moore machine can have a size (represented by its number of states) that is far less than the density of the sparsest winning positional strategy in a game. For example, a game with many positions could be winning for the system player by always playing the same action. A machine representing this strategy would only have one state, whereas many positions of the safety player in the game might be visited along a corresponding play. While it is always possible to translate a winning positional strategy of some density $n$ into a Mealy machine of size at most $n+1$ (assuming that player $1$ plays first in the game), the DFA produced by a computational learning approach is equally suitable as a starting point for a Mealy machine computation: we use the state set of the DFA as state set of the Mealy machine, but contract a sequence of two successive transitions that represents the input and the output in one round to one transition in the Mealy machine. The number of states that then remain reachable is the size of the Mealy machine. Figure \ref{fig:makingMealyMachine} illustrates this translation process. For a more thorough definition and discussion of the connection between Mealy/Moore machines and games, see \cite{DBLP:journals/fmsd/Ehlers12}.

\section{Benchmarks}
\label{sec:benchmarks}
\label{sec:BoundedSynthesis}

To make our experimental evaluation as insightful as possible, we only consider games from practice as benchmarks, and leave out the commonly used randomly generated games and toy examples such as variants of tic-tac-toe or other folk games. Instead, we use games stemming from \newterm{Bounded Synthesis} \cite{DBLP:conf/atva/ScheweF07a}, which is an approach for the synthesis of finite-state systems from specifications in temporal logic. Intuitively, a synthesis process that follows this approach starts by representing the specification as an automaton that ensures that for every trace of a system to be synthesized that we declare to be illegal, the automaton has some corresponding run on which some so-called rejecting state is visited infinitely often. If we now restrict the number of visits to these rejecting states along a run to some finite value $b$, and find a system for which all automaton runs for all traces of the system visit the rejecting states of the automaton only at most $b$ times, then we have a valid implementation. At the same time, the problem of synthesizing such solutions can be reduced to safety game solving, which makes the approach conceptually simple.

Here, we consider two variants of building the games from specifications. The first one uses the classical construction from \cite{DBLP:conf/atva/ScheweF07a}, adapted to finding Mealy machines instead of Moore machine implementations. In the second one, we use a modification proposed in \cite{DBLP:journals/fmsd/Ehlers12}: we allow the system player to voluntarily put herself into an unnecessarily bad situation in the game. In a bounded synthesis game, positions are labelled by some counter vector $(c_1, \ldots, c_n)$, which are updated whenever both players have made their moves. The positions have the property that for two positions $v$ and $v'$ labelled by $(c_1, \ldots, c_n)$ and $(c'_1, \ldots, c'_n)$ such that for every $i \in \{1, \ldots, n\}$, we have $c_i \geq c'_i$, all of player 0's winning strategies for $v^\mathit{init} = v$ are also winning strategies for the same game but with $v^\mathit{init} = v'$. Thus, by allowing player $0$ to increase her counters voluntarily, we do not give her additional power. Additionally, we introduce a position for player $0$ to increase her counter values from the initial ones before the actual start of the game. While this modification does not give player $0$ more possibilities to win the game, it allows us to find sparser strategies. In fact, it is a corollary of Theorem 2 of \cite{DBLP:conf/atva/ScheweF07a} that if and only if there exists some Mealy machine with $n$ states that satisfies the specification and adheres to a bound of $b$, then the bounded synthesis game with the counter increase possibility for player $0$ will allow a strategy of density $n \cdot |\Sigma^1| + 1$. Thus, searching for the sparsest positional strategy will lead to the smallest Mealy-type implementation. 
Note that strictly speaking, the safety games resulting from the modification do not conform to the safety game definition in Section \ref{sec:definitions} any more, as the counter increasing possibility leads to multiple successors that all correspond to the same action for some positions of player $0$. However, for  approaches to find sparse positional winning strategies, this makes no difference. 
For both variants of Bounded Synthesis, we consider the following benchmarks:
\begin{itemize}
 \item a basic mutex ($\mathit{BasicMutex}$), for the linear-time temporal logic (LTL) specification $\psi = \LTLG ( r_1 \rightarrow \LTLF g_1) \wedge \LTLG ( r_2 \rightarrow \LTLF g_2) \wedge \LTLG ( \neg g_1 \vee \neg g_2)$, the input bits $\{r_1,r_2\}$, and the output bits $\{g_1,g_2\}$,
 \item a basic reaction scheme ($\mathit{BasicReaction}$) with the specification $\psi = (x \rightarrow \LTLG \neg z) \wedge (\neg x \rightarrow \LTLG z)$ for the input bits $\{x\}$ and the output bits $\{z\}$,
 \item three dining philosophers ($\mathit{ThreePhilosophers}$) getting hungry at the same time, with $\psi = \LTLG (h \rightarrow \LTLX (\LTLF e_1 \wedge \LTLF e_2 \wedge \LTLF e_3)) \wedge \LTLG( (\neg e_1 \vee \neg e_2) \wedge (\neg e_1 \vee \neg e_3) \wedge (\neg e_2 \vee \neg e_3))$ for the input bit $h$ and the output bits $\{e_1,e_2,e_3\}$ (describing which philosophers are eating), and
 \item some examples from \cite{DBLP:conf/fmcad/JobstmannB06}, mostly arbiter and traffic light examples ($\mathit{demo-v3}\ \allowbreak \ldots\ \allowbreak \mathit{demo-v23}$). Unrealizable specifications have been left out.
\end{itemize}
All benchmarks are parametrized by the bound value. For example, the table entry $\mathit{BasicMutex}(3)$ in the following section will refer to the basic mutex example with a bound value of $3$. In the case that the second variant of the Bounded Synthesis process is used, in which player $0$ has the possibility to voluntarily increase some counter values, the benchmark name appears primed, e.g., as in $\mathit{BasicMutex}'(3)$.

\section{Experimental Results}
\label{sec:experiments}

We implemented the approaches described in Section \ref{sec:approaches} in C++, except for the learning approach \cite{DBLP:conf/atva/Neider11}, for which we use an implementation provided by the author of \cite{DBLP:conf/atva/Neider11} (also written in C++). For \textsc{OPTSAT} \cite{DBLP:conf/jelia/GiunchigliaM06} and the learning-based tool, we used default settings. As (integer) linear programming library, we took \textsc{liblpsolve v.5.5}.

For obtaining the benchmarks, we implemented a tool that computes safety games for the Bounded Synthesis approach, without using any symbolic data structure such as binary decision diagrams (BDDs). Benchmarks for which the preparation required more than 64 gigabytes of RAM were left out. This limit was frequently exceeded for the modified Bounded Synthesis approach, as many of the resulting games have a huge number of positions, even though the percentage of positions that are winning for the safety player, and thus are input to the strategy density optimization algorithms, is quite low. Typically, we scaled the bound for the synthesis benchmarks up to $5$. If the number of winning positions in a game exceeds 10000 for some bound $b$, or if increasing the bound would yield the same game, we did however not consider higher bounds. All games were pruned to the positions reachable when player $0$ follows some arbitrary specialization of the most permissive strategy. 

We used a Sun XFire computer with 2.6 Ghz AMD Opteron processors running an x64-version of Linux for obtaining the results. All tools considered are single-threaded. We restricted the memory
usage for the strategy extraction to 4 GB and set a timeout of 600 seconds per invocation. All tools were ran five times (25 times for the randomized approaches) to level out fluctuations. The tables in the following represent mean values.

\subsection{Strategy Densities}

Table \ref{tab:sizeBoundedSynthesis} and Table \ref{tab:sizeBoundedSynthesis2} compare the obtained strategy densities (or sizes for non-positional strategies) on the classical Bounded Synthesis games, whereas Table \ref{tab:sizeBoundedSynthesisMod} considers the Bounded Synthesis benchmarks with the modification that player $0$ can increase counter values at will. Timeouts are represented by ``t/o''. Since for the modification switched on, building the safety games resulted in running out of memory in many cases, Table \ref{tab:sizeBoundedSynthesisMod} only has relatively few entries. The remaining benchmarks have a low to medium number of positions, as the non-winning positions have already been pruned away, and these constitute the majority of positions created while building the game. However, the large search spaces and the bad performance of the purely random strategy extraction approach show that the benchmarks are still far from being trivial. The search space size (in bits) represents how many syntactically different positional strategies are possible, and is defined to be $\sum_{v \in V_0}\log_2( \allowbreak |\{ x \in \Sigma^0 \mid E(v,x) \in W|)$ for the set of winning positions $W$. Quite often, the sparsest strategies only have a very low density. This is not a surprising situation in synthesis, as many systems can be implemented in very few states. 
The combination of large search spaces and the availability of sparse winning strategies make the benchmarks at hand an excellent competition ground for the sparse strategy extraction approaches. To compare the density of positional strategies and the size of learning-based strategies, for all tables, the number of input and output atomic propositions in the benchmark are also given. 

It can be seen that for both Bounded Synthesis variants, the randomized approach and the repetitive linear-programming approach are quite competitive against the exact minimization approaches (ILP and OPTSAT), despite the large search space. For many benchmarks for which very sparse strategies are possible (e.g., $\mathit{demo-v8},\mathit{demo-v12},\mathit{demo-v13}$), all of the approaches dealing with positional strategies find some sparsest strategy. Furthermore, there is no clear winner of the smart randomized approach and repetitive linear programming. For example, for the basic mutex (unprimed), the latter approach always finds a sparsest strategy, whereas the randomized approach does not. On the other hand, for the dining philosophers (unprimed) and other benchmarks like $\mathit{demo-v9}$, the situation is reversed.

When evaluating how well the computational learning approach works, we need to compare across tables. In Table \ref{tab:sizeBoundedSynthesisMod}, the approach is not listed. The reason is that due to the counter increase option of player $0$, there can be many successors in the game that all correspond to the same action, and the implementation of the approach does not support such games. However, since the learning approach can already find the smallest strategy in the games produced in the classical Bounded Synthesis approach, this is no drawback. Recall from Section \ref{sec:benchmarks} that if and only if there exists a Mealy machine with $n$ states that satisfies the specification and respects some bound $b$, then in the modified synthesis game for bound $b$, there is a positional strategy with density $n \cdot |\Sigma^1|+1$. This fact allows us to measure the success of the learning approach. We can see than in most cases, it did not find the minimal implementation. For example, for $\mathit{BasicMutex'(1)}$, the sparsest positional strategy has density $9$, i.e., $n \cdot |\Sigma^1| + 1$ for $n= 2$. Intuitively, for this benchmark, a Mealy machine with two states that satisfies the specification would simply alternate between giving the grant to the two requesters. The Mealy machine sizes for the learning approach and $\mathit{BasicMutex(b)}$ for $b = \{1, \ldots, 5\}$ are however larger, and grow with the values of $b$. Thus, the learning approach can be fooled by needlessly large games. However, for benchmarks such as $\mathit{demo-v22}(1)$, for which the modified version of the game was too large to fit into 64 GB of memory, the learning approach can deal well with the classical version of the game: a Mealy machine with 13 states is found, although the sparsest positional strategy has $201$ reachable positions of $V^0$, for $|\Sigma^1|=8$. The benchmark $\mathit{demo-v22}$ represents an elevator controller synthesis problem. For comparison, the numbers of states of the deterministic finite automata produced from the benchmarks in the learning-based approach are also given in Table \ref{tab:sizeBoundedSynthesis} and Table  \ref{tab:sizeBoundedSynthesis2}.

\begin{table}
\caption{Strategy density/size comparison for the classical Bounded Synthesis benchmarks (part one). }
\label{tab:sizeBoundedSynthesis}
\scriptsize\centering
\renewcommand{\tabcolsep}{1mm} \begin{tabular}{l||c|c|c|c|c||c|c|c|c|c|c|c} \textbf{Benchmark} & \textbf{$|V^0|$} & \textbf{$|V^1|$} &  \textbf{$|\Sigma^0|$} & \textbf{$|\Sigma^1|$} & \textbf{Search space} & \textbf{Random} & \textbf{ILP} & \textbf{OPTSAT} & \textbf{Random} & \textbf{RepLP} & \textbf{Learning} & \textbf{Learning} \\ & & & & & \textbf{(bits)} & \textbf{(dumb)} & & & \textbf{(smart)} & & \textbf{(aut.)} & \textbf{(Mealy)} \\ \hline
$\mathit{BasicMutex}(1)$  & 13  & 3  & 4  & 4  & 3  & 13  & 13  & 13  & 13  & 13  & 10  & 3  \\ \hline 
$\mathit{BasicMutex}(2)$  & 33  & 8  & 4  & 4  & 17.9248  & 27.4  & 13  & 13  & 14.12  & 13  & 20  & 7  \\ \hline 
$\mathit{BasicMutex}(3)$  & 61  & 15  & 4  & 4  & 47.2842  & 43.56  & 13  & 13  & 16.52  & 13  & 34  & 13  \\ \hline 
$\mathit{BasicMutex}(4)$  & 97  & 24  & 4  & 4  & 89.3233  & 66.6  & 13  & 13  & 17  & 13  & 52  & 21  \\ \hline 
$\mathit{BasicMutex}(5)$  & 141  & 35  & 4  & 4  & 144.042  & 86.6  & 13  & 13  & 17  & 13  & 74  & 31  \\ \hline 
$\mathit{BasicReaction}(1)$  & 7  & 3  & 2  & 2  & 0  & 7  & 7  & 7  & 7  & 7  & 5  & 3  \\ \hline 
$\mathit{ThreePhilosophers}(3)$  & 53  & 26  & 8  & 2  & 15.1699  & 51.08  & 41  & 41  & 41  & 43.2  & 30  & 11  \\ \hline 
$\mathit{ThreePhilosophers}(4)$  & 151  & 75  & 8  & 2  & 107.699  & 117.8  & 37  & 37  & 41.16  & 43.6  & 50  & 19  \\ \hline 
$\mathit{ThreePhilosophers}(5)$  & 309  & 154  & 8  & 2  & 291.248  & 211.2  & 35  & 35  & 48.36  & 59.2  & 74  & 29  \\ \hline 
$\mathit{demo-v3}(2)$  & 113  & 14  & 2  & 8  & 32  & 113  & 113  & 113  & 113  & 113  & 26  & 10  \\ \hline 
$\mathit{demo-v3}(3)$  & 177  & 22  & 2  & 8  & 56  & 169.6  & 113  & 113  & 113  & 113  & 29  & 12  \\ \hline 
$\mathit{demo-v4}(3)$  & 209  & 26  & 2  & 8  & 56  & 206.4  & 177  & 177  & 177  & 177  & 36  & 15  \\ \hline 
$\mathit{demo-v5}(2)$  & 145  & 18  & 2  & 8  & 60  & 135.4  & 113  & 113  & 113  & 113.4  & 27  & 11  \\ \hline 
$\mathit{demo-v5}(3)$  & 193  & 24  & 2  & 8  & 104  & 173.8  & 113  & 113  & 113  & 114  & 29  & 12 \\ \hline 
$\mathit{demo-v6}(3)$  & 321  & 40  & 2  & 8  & 168  & 275.6  & 177  & 177  & 177  & 177.2  & 37  & 16  \\ \hline 
$\mathit{demo-v7}(1)$  & 81  & 10  & 2  & 8  & 34  & 80.68  & 65  & 65  & 65  & 66  & 17  & 7  \\ \hline 
$\mathit{demo-v7}(2)$  & 105  & 13  & 2  & 8  & 56  & 98.6  & 65  & 65  & 65  & 65.4  & 19  & 8  \\ \hline 
$\mathit{demo-v7}(3)$  & 137  & 17  & 2  & 8  & 84  & 122.9  & 65  & 65  & 65  & 66  & 21  & 9  \\ \hline 
$\mathit{demo-v8}(1)$  & 9  & 4  & 2  & 2  & 6  & 5.32  & 3  & 3  & 3  & 3  & 6  & 3  \\ \hline 
$\mathit{demo-v8}(2)$  & 13  & 6  & 2  & 2  & 10  & 6.2  & 3  & 3  & 3  & 3  & 8  & 4  \\ \hline 
$\mathit{demo-v8}(3)$  & 17  & 8  & 2  & 2  & 14  & 6.12  & 3  & 3  & 3  & 3  & 10  & 5  \\ \hline 
$\mathit{demo-v8}(4)$  & 21  & 10  & 2  & 2  & 18  & 6.36  & 3  & 3  & 3  & 3  & 12  & 6  \\ \hline 
$\mathit{demo-v8}(5)$  & 25  & 12  & 2  & 2  & 22  & 6.52  & 3  & 3  & 3  & 3  & 14  & 7  \\ \hline 
$\mathit{demo-v9}(1)$  & 13  & 6  & 2  & 2  & 6  & 11.32  & 9  & 9  & 9  & 9.6  & 9  & 4  \\ \hline 
$\mathit{demo-v9}(2)$  & 17  & 8  & 2  & 2  & 9  & 13.72  & 9  & 9  & 9  & 9.2  & 18  & 8  \\ \hline 
$\mathit{demo-v9}(3)$  & 21  & 10  & 2  & 2  & 12  & 12.68  & 9  & 9  & 9  & 9.4  & 23  & 10  \\ \hline 
$\mathit{demo-v9}(4)$  & 25  & 12  & 2  & 2  & 15  & 12.76  & 9  & 9  & 9  & 10  & 28  & 12  \\ \hline 
$\mathit{demo-v9}(5)$  & 29  & 14  & 2  & 2  & 18  & 14.6  & 9  & 9  & 9  & 10  & 33  & 14  \\ \hline 
$\mathit{demo-v10}(1)$  & 53  & 13  & 4  & 4  & 56  & 38.12  & 13  & 13  & 13  & 13  & 10  & 5  \\ \hline 
$\mathit{demo-v10}(2)$  & 89  & 22  & 4  & 4  & 80  & 51.08  & 13  & 13  & 13  & 13  & 16  & 8  \\ \hline 
$\mathit{demo-v10}(3)$  & 141  & 35  & 4  & 4  & 120  & 65.96  & 13  & 13  & 13  & 13  & 24  & 12  \\ \hline 
$\mathit{demo-v10}(4)$  & 209  & 52  & 4  & 4  & 176  & 85.64  & 13  & 13  & 13  & 13  & 34  & 17  \\ \hline 
$\mathit{demo-v10}(5)$  & 293  & 73  & 4  & 4  & 248  & 134  & 13  & 13  & 13  & 13  & 46  & 23  \\ \hline 
$\mathit{demo-v12}(1)$  & 33  & 8  & 4  & 4  & 50.3399  & 26.6  & 9  & 9  & 9  & 9  & 2  & 1  \\ \hline 
$\mathit{demo-v12}(2)$  & 37  & 9  & 4  & 4  & 58.3399  & 27.72  & 9  & 9  & 9  & 9  & 2  & 1  \\ \hline 
$\mathit{demo-v12}(3)$  & 41  & 10  & 4  & 4  & 66.3399  & 30.44  & 9  & 9  & 9  & 9  & 2  & 1  \\ \hline 
$\mathit{demo-v12}(4)$  & 45  & 11  & 4  & 4  & 74.3399  & 32.84  & 9  & 9  & 9  & 9  & 2  & 1  \\ \hline 
$\mathit{demo-v12}(5)$  & 49  & 12  & 4  & 4  & 82.3399  & 29.32  & 9  & 9  & 9  & 9  & 2  & 1  \\ \hline 
$\mathit{demo-v13}(1)$  & 5  & 2  & 2  & 2  & 0  & 5  & 5  & 5  & 5  & 5  & 4  & 2  \\ \hline 
$\mathit{demo-v13}(2)$  & 9  & 4  & 2  & 2  & 2  & 6.6  & 5  & 5  & 5  & 5  & 8  & 4  \\ \hline 
$\mathit{demo-v13}(3)$  & 13  & 6  & 2  & 2  & 4  & 8.68  & 5  & 5  & 5  & 5  & 12  & 6  \\ \hline 
$\mathit{demo-v13}(4)$  & 17  & 8  & 2  & 2  & 6  & 6.92  & 5  & 5  & 5  & 5  & 16  & 8  \\ \hline 
$\mathit{demo-v13}(5)$  & 21  & 10  & 2  & 2  & 8  & 8.92  & 5  & 5  & 5  & 5  & 20  & 10  \\ \hline 
$\mathit{demo-v14}(1)$  & 49  & 12  & 4  & 4  & 47.794  & 43.08  & 13  & 13  & 13  & 13  & 15  & 8  \\ \hline 
$\mathit{demo-v14}(2)$  & 129  & 32  & 4  & 4  & 145.137  & 98.44  & 13  & 13  & 13  & 13  & 29  & 15  \\ \hline 
$\mathit{demo-v14}(3)$  & 241  & 60  & 4  & 4  & 297.293  & 154.1  & 13  & 13  & 13  & 13  & 47  & 24  \\ \hline 
$\mathit{demo-v14}(4)$  & 385  & 96  & 4  & 4  & 500.168  & 243.6  & 13  & 13  & 13  & 13  & 69  & 35  \\ \hline 
$\mathit{demo-v14}(5)$  & 561  & 140  & 4  & 4  & 753.762  & 350  & 13  & 13  & 13  & 13  & 95  & 48  \\ \hline 
$\mathit{demo-v15}(1)$  & 37  & 9  & 4  & 4  & 15  & 35.56  & 29  & 29  & 29  & 29  & 13  & 4  \\ \hline 
$\mathit{demo-v15}(2)$  & 65  & 16  & 4  & 4  & 42.6045  & 57  & 29  & 29  & 31.4  & 29.8  & 25  & 9  \\ \hline 
$\mathit{demo-v15}(3)$  & 101  & 25  & 4  & 4  & 82.3038  & 86.28  & 29  & 29  & 32.68  & 29  & 41  & 16  \\ \hline 
$\mathit{demo-v15}(4)$  & 145  & 36  & 4  & 4  & 134.683  & 111.1  & 29  & 29  & 34.6  & 29.2  & 61  & 25  \\ \hline 
$\mathit{demo-v15}(5)$  & 197  & 49  & 4  & 4  & 199.741  & 138.6  & 29  & 29  & 35.56  & 30.6  & 85  & 36 \\ \hline
$\mathit{demo-v16}(2)$  & 409  & 51  & 8  & 8  & 270.176  & 385.6  & 217  & 217  & 228.2  & 217.8  & 59  & 17  \\ \hline 
$\mathit{demo-v16}(3)$  & 873  & 109  & 8  & 8  & 764.786  & 783.7  & 209  & 209  & 255.1  & 219  & 138  & 47  \\ \hline 
$\mathit{demo-v16}(4)$  & 1577  & 197  & 8  & 8  & 1623.06  & 1367  & 209  & t/o  & 290.6  & 221  & 252  & 93  \\ \hline 
$\mathit{demo-v16}(5)$  & 2569  & 321  & 8  & 8  & 2941.41  & 2109  & t/o  & t/o  & 322.6  & 217.8  & 420  & 164  \\ \hline 
$\mathit{demo-v17}(2)$  & 209  & 52  & 8  & 4  & 188.229  & 168.7  & 25  & 25  & 29.48  & 25  & 51  & 26
\end{tabular}\end{table}

\begin{table}
\caption{Strategy density/size comparison for the classical Bounded Synthesis benchmarks (part two).}
\label{tab:sizeBoundedSynthesis2}
\scriptsize\centering
\renewcommand{\tabcolsep}{1mm}\begin{tabular}{l||c|c|c|c|c||c|c|c|c|c|c|c} \textbf{Benchmark} & \textbf{$|V^0|$} & \textbf{$|V^1|$} &  \textbf{$|\Sigma^0|$} & \textbf{$|\Sigma^1|$} & \textbf{Search space} & \textbf{Random} & \textbf{ILP} & \textbf{OPTSAT} & \textbf{Random} & \textbf{RepLP} & \textbf{Learning} & \textbf{Learning} \\ & & & & & \textbf{(bits)} & \textbf{(dumb)} & & & \textbf{(smart)} & & \textbf{(aut.)} & \textbf{(Mealy)} \\ \hline
$\mathit{demo-v17}(3)$  & 733  & 183  & 8  & 4  & 863.098  & 526.9  & 25  & 25  & 40.04  & 26.6  & 139  & 70  \\ \hline 
$\mathit{demo-v17}(4)$  & 1641  & 410  & 8  & 4  & 2187.1  & 1084  & 25  & t/o  & 59.88  & 26.6  & 273  & 137  \\ \hline 
$\mathit{demo-v17}(5)$  & 3029  & 757  & 8  & 4  & 4343.1  & 1851  & 25  & t/o  & 51.24  & 26.2  & 491  & 246  \\ \hline 
$\mathit{demo-v18}(3)$  & 6273  & 784  & 16  & 8  & 7501.28  & 5477  & t/o  & t/o  & 338.3  & 253.8  & 439  & 218  \\ \hline 
$\mathit{demo-v19}(1)$  & 65  & 16  & 4  & 4  & 53.5489  & 53.8  & 17  & 17  & 21.32  & 17  & 20  & 7  \\ \hline 
$\mathit{demo-v19}(2)$  & 125  & 31  & 4  & 4  & 126.702  & 93.64  & 17  & 17  & 25.8  & 17.2  & 38  & 13  \\ \hline 
$\mathit{demo-v19}(3)$  & 201  & 50  & 4  & 4  & 226.385  & 135.9  & 17  & 17  & 27.24  & 17  & 61  & 21  \\ \hline 
$\mathit{demo-v19}(4)$  & 293  & 73  & 4  & 4  & 351.427  & 184  & 17  & 17  & 35.4  & 17  & 90  & 31  \\ \hline 
$\mathit{demo-v19}(5)$  & 401  & 100  & 4  & 4  & 501.829  & 218.6  & 17  & 17  & 32.2  & 17  & 125  & 43  \\ \hline 
$\mathit{demo-v20}(1)$  & 509  & 127  & 8  & 4  & 741.196  & 266.4  & 17  & 17  & 29.48  & 21  & 4  & 3  \\ \hline 
$\mathit{demo-v20}(2)$  & 1585  & 396  & 8  & 4  & 2755.88  & 404.4  & 17  & 17  & 31.72  & 21.4  & 6  & 4  \\ \hline 
$\mathit{demo-v20}(3)$  & 3081  & 770  & 8  & 4  & 5696.31  & 701  & 17  & 17  & 37.8  & 21.2  & 8  & 5  \\ \hline 
$\mathit{demo-v20}(4)$  & 5217  & 1304  & 8  & 4  & 9754.35  & 1120  & 17  & t/o  & 41.32  & 21.2  & 10  & 6  \\ \hline 
$\mathit{demo-v21}(1)$  & 97  & 6  & 16  & 16  & 0  & 97  & 97  & 97  & 97  & 97  & 31  & 6 \\ \hline 
$\mathit{demo-v21}(2)$  & 417  & 26  & 16  & 16  & 20  & 417  & 97  & 97  & 97  & 97  & 111  & 13  \\ \hline 
$\mathit{demo-v21}(3)$  & 2017  & 126  & 16  & 16  & 155.098  & 2017  & 97  & 97  & 97  & 97  & 333  & 74  \\ \hline 
$\mathit{demo-v22}(1)$  & 353  & 44  & 2  & 8  & 40  & 314.3  & 201  & 201  & 201  & 201  & 28  & 13  \\ \hline 
$\mathit{demo-v22}(2)$  & 505  & 63  & 2  & 8  & 139  & 454.4  & 201  & 201  & 227.9  & 201  & 37  & 18  \\ \hline 
$\mathit{demo-v22}(3)$  & 633  & 79  & 2  & 8  & 211  & 582.1  & 201  & 201  & 229.5  & 201.4  & 47  & 23  \\ \hline 
$\mathit{demo-v22}(4)$  & 761  & 95  & 2  & 8  & 268  & 703.1  & 201  & 201  & 228.5  & 202.8  & 57  & 28  \\ \hline 
$\mathit{demo-v22}(5)$  & 889  & 111  & 2  & 8  & 325  & 816.4  & 201  & 201  & 216.4  & 204  & 67  & 33  \\ \hline 
$\mathit{demo-v23}(1)$  & 27  & 13  & 2  & 2  & 11  & 20.44  & 15  & 15  & 15.56  & 15  & 10  & 5  \end{tabular}\end{table}

\begin{table}
\caption{Strategy density comparison for the Bounded Synthesis benchmarks, with modification to allow for sparser strategies.}
\label{tab:sizeBoundedSynthesisMod}
\scriptsize\centering
\renewcommand{\tabcolsep}{1mm} \begin{tabular}{l||c|c|c|c|c||c|c|c|c|c} \textbf{Benchmark} & \textbf{$|V^0|$} & \textbf{$|V^1|$} & \textbf{$|\Sigma^0|$} & \textbf{$|\Sigma^1|$} & \textbf{Search space} & \textbf{Random} & \textbf{ILP} & \textbf{OPTSAT} & \textbf{Random} & \textbf{RepLP} \\ & & & & & \textbf{(bits)} & \textbf{(dumb)} & & & \textbf{(smart)} &   \\ \hline
$\mathit{BasicMutex}'(1)$  & 33  & 8  & 4  & 4  & 66  & 28.04  & 9  & 9  & 9  & 9  \\ \hline 
$\mathit{BasicMutex}'(2)$  & 121  & 30  & 4  & 4  & 414.762  & 60.04  & 9  & 9  & 9  & 9  \\ \hline 
$\mathit{BasicMutex}'(3)$  & 289  & 72  & 4  & 4  & 1298.79  & 130.8  & 9  & 9  & 9  & 9  \\ \hline 
$\mathit{BasicMutex}'(4)$  & 561  & 140  & 4  & 4  & 2998.44  & 214.6  & 9  & 9  & 9  & 9  \\ \hline 
$\mathit{BasicMutex}'(5)$  & 961  & 240  & 4  & 4  & 5816.8  & 348.4  & 9  & 9  & 9  & 9  \\ \hline 
$\mathit{BasicReaction}'(1)$  & 13  & 6  & 2  & 2  & 7  & 7.88  & 7  & 7  & 7  & 7.2  \\ \hline 
$\mathit{BasicReaction}'(2)$  & 19  & 9  & 2  & 2  & 17.0947  & 7.96  & 7  & 7  & 7  & 8  \\ \hline 
$\mathit{BasicReaction}'(3)$  & 25  & 12  & 2  & 2  & 29.5098  & 7.88  & 7  & 7  & 7  & 7.2  \\ \hline 
$\mathit{BasicReaction}'(4)$  & 31  & 15  & 2  & 2  & 43.7633  & 9.64  & 7  & 7  & 7  & 7.4  \\ \hline 
$\mathit{BasicReaction}'(5)$  & 37  & 18  & 2  & 2  & 59.5361  & 9.24  & 7  & 7  & 7  & 8  \\ \hline 
$\mathit{ThreePhilosophers}'(3)$  & 217  & 108  & 8  & 2  & 865.396  & 53.4  & 7  & 7  & 7  & 7  \\ \hline 
$\mathit{ThreePhilosophers}'(4)$  & 785  & 392  & 8  & 2  & 4246.45  & 83.16  & t/o  & 7  & 7  & 7  \\ \hline 
$\mathit{ThreePhilosophers}'(5)$  & 1921  & 960  & 8  & 2  & 12504.8  & 114.3  & t/o  & 7  & 7  & 7  \\ \hline 
$\mathit{demo-v8}'(1)$  & 37  & 18  & 2  & 2  & 136.287  & 16.44  & 3  & 3  & 3  & 3  \\ \hline 
$\mathit{demo-v8}'(2)$  & 97  & 48  & 2  & 2  & 472.304  & 31.24  & 3  & 3  & 3  & 3  \\ \hline 
$\mathit{demo-v8}'(3)$  & 201  & 100  & 2  & 2  & 1164.66  & 56.92  & 3  & 3  & 3  & 3  \\ \hline 
$\mathit{demo-v8}'(4)$  & 361  & 180  & 2  & 2  & 2365.96  & 83.64  & 3  & 3  & 3  & 3  \\ \hline 
$\mathit{demo-v8}'(5)$  & 589  & 294  & 2  & 2  & 4239.85  & 118.5  & 3  & 3  & 3  & 3  \\ \hline 
$\mathit{demo-v9}'(1)$  & 649  & 324  & 2  & 2  & 2517.11  & 39.32  & 5  & 5  & 5.24  & 6  \\ \hline 
$\mathit{demo-v9}'(2)$  & 4609  & 2304  & 2  & 2  & 24739  & 71.32  & t/o  & 5  & 5.56  & 6.2  \\ \hline 
$\mathit{demo-v13}'(1)$  & 55  & 27  & 2  & 2  & 259.934  & 42.2  & 3  & 3  & 3  & 3  \\ \hline 
$\mathit{demo-v13}'(2)$  & 129  & 64  & 2  & 2  & 773  & 105.8  & 3  & 3  & 3  & 3  \\ \hline 
$\mathit{demo-v13}'(3)$  & 251  & 125  & 2  & 2  & 1747.67  & 196.9  & 3  & 3  & 3  & 3  \\ \hline 
$\mathit{demo-v13}'(4)$  & 433  & 216  & 2  & 2  & 3357.28  & 347  & 3  & 3  & 3  & 3  \\ \hline 
$\mathit{demo-v13}'(5)$  & 687  & 343  & 2  & 2  & 5785.47  & 547.3  & 3  & 3  & 3  & 3  \\ \hline 
$\mathit{demo-v15}'(1)$  & 169  & 42  & 4  & 4  & 542.836  & 104.2  & 13  & 13  & 13.96  & 14.6  \\ \hline 
$\mathit{demo-v15}'(2)$  & 769  & 192  & 4  & 4  & 3614.23  & 261  & 13  & 13  & 14.28  & 14.6  \\ \hline 
$\mathit{demo-v19}'(1)$  & 2241  & 560  & 4  & 4  & 12946.8  & 781.5  & 9  & 9  & 9.8  & 9  \\ \hline 
$\mathit{demo-v23}'(1)$  & 2521  & 1260  & 2  & 2  & 14742  & 186.8  & t/o  & 5  & 6.84  & 5  \end{tabular}\end{table}

\subsection{Computation Times}

Table \ref{table:computationTimeClassic} presents computation times for the classical Bounded Synthesis benchmarks, whereas Table \ref{table:computationTimeModified} describes the results for the modified version. For brevity, benchmarks for which all tools needed less than 50\,ms of computation times have been left out.

The tables show no big surprises. The exact approaches time out for the largest benchmarks. For the benchmarks stemming from the modified Bounded Synthesis version, \textsc{OPTSAT} performs better than the ILP-based approach, whereas for the non-modified version, the ILP solver seems to be faster. The main difference between the two classes is the fact that the number of successors of positions of player $0$ is much higher in the modified synthesis games. OPTSAT seems to be able to deal with this situation in a better way.
The learning approach is typically slower than the heuristics for positional strategies, but unlike the exact approaches, did not time out for any of the benchmarks.

\begin{table}
\caption{Computation time comparison for the classical Bounded Synthesis benchmarks. All times are given in seconds.}
\label{table:computationTimeClassic}
\scriptsize\centering
\begin{tabular}{l||c|c|c|c|c|c} \textbf{Benchmark} & \textbf{Random} & \textbf{ILP} & \textbf{OPTSAT} & \textbf{Random} & \textbf{RepLP} & \textbf{Learning}  \\ & \textbf{(dumb)} & & & \textbf{(smart)} & & \\ \hline
$\mathit{BasicMutex}(4)$  & 0.00671  & 0.00948  & 0.06751  & 0.00804  & 0.0204  & 0.038  \\ \hline 
$\mathit{BasicMutex}(5)$  & 0.00865  & 0.0109  & 0.4895  & 0.00736  & 0.0107  & 0.07821  \\ \hline 
$\mathit{ThreePhilosophers}(4)$  & 0.00795  & 2.58  & 3.241  & 0.00847  & 0.059  & 0.04376  \\ \hline 
$\mathit{ThreePhilosophers}(5)$  & 0.00925  & 129  & 89.7  & 0.0114  & 0.241  & 0.114  \\ \hline 
$\mathit{demo-v3}(2)$  & 0.00659  & 0.0132  & 0.0641  & 0.0117  & 0.0262  & 0.02546  \\ \hline 
$\mathit{demo-v6}(3)$  & 0.00874  & 0.0427  & 0.04663  & 0.0335  & 0.0584  & 0.06382  \\ \hline 
$\mathit{demo-v10}(2)$  & 0.00629  & 0.00974  & 0.07321  & 0.00747  & 0.00999  & 0.01083  \\ \hline 
$\mathit{demo-v10}(5)$  & 0.00831  & 0.0165  & 0.1267  & 0.00979  & 0.0151  & 0.07951  \\ \hline 
$\mathit{demo-v14}(2)$  & 0.00675  & 0.0133  & 0.05853  & 0.00664  & 0.0123  & 0.03176  \\ \hline 
$\mathit{demo-v14}(3)$  & 0.00845  & 0.0174  & 0.04967  & 0.0104  & 0.0156  & 0.1244  \\ \hline 
$\mathit{demo-v14}(4)$  & 0.00896  & 0.0237  & 0.3457  & 0.0112  & 0.0189  & 0.4291  \\ \hline 
$\mathit{demo-v14}(5)$  & 0.0107  & 0.0337  & 4.473  & 0.0134  & 0.0261  & 1.271  \\ \hline 
$\mathit{demo-v15}(4)$  & 0.00677  & 0.0177  & 0.1765  & 0.00989  & 0.0167  & 0.0701  \\ \hline 
$\mathit{demo-v15}(5)$  & 0.00834  & 0.0266  & 0.1027  & 0.00938  & 0.0187  & 0.1399  \\ \hline 
$\mathit{demo-v16}(2)$  & 0.00903  & 0.408  & 0.2805  & 0.035  & 0.555  & 0.1248  \\ \hline 
$\mathit{demo-v16}(3)$  & 0.0144  & 84.9  & 86.47  & 0.0589  & 1.46  & 0.846  \\ \hline 
$\mathit{demo-v16}(4)$  & 0.0219  & 548  & t/o  & 0.111  & 3.24  & 5.05  \\ \hline 
$\mathit{demo-v16}(5)$  & 0.0325  & t/o  & t/o  & 0.214  & 8.11  & 19.91  \\ \hline 
$\mathit{demo-v17}(2)$  & 0.00762  & 0.0345  & 0.1598  & 0.00947  & 0.0205  & 0.1075  \\ \hline 
$\mathit{demo-v17}(3)$  & 0.0123  & 0.509  & 16.84  & 0.017  & 0.0906  & 1.963  \\ \hline 
$\mathit{demo-v17}(4)$  & 0.0245  & 2.02  & t/o  & 0.0455  & 0.253  & 20.97  \\ \hline 
$\mathit{demo-v17}(5)$  & 0.0353  & 4.45  & t/o  & 0.0691  & 0.571  & 163  \\ \hline 
$\mathit{demo-v18}(3)$  & 0.0885  & t/o  & t/o  & 0.917  & 92.8  & 247.5  \\ \hline 
$\mathit{demo-v19}(3)$  & 0.00761  & 0.0142  & 0.2191  & 0.00896  & 0.0156  & 0.06583  \\ \hline 
$\mathit{demo-v19}(4)$  & 0.00764  & 0.0173  & 1.272  & 0.0106  & 0.0193  & 0.1379  \\ \hline 
$\mathit{demo-v19}(5)$  & 0.00929  & 0.0219  & 1.51  & 0.0127  & 0.022  & 0.2704  \\ \hline 
$\mathit{demo-v20}(1)$  & 0.0127  & 0.0323  & 0.9498  & 0.0131  & 0.0377  & 0.01549  \\ \hline 
$\mathit{demo-v20}(2)$  & 0.0206  & 0.222  & 15.81  & 0.0288  & 0.197  & 0.06499  \\ \hline 
$\mathit{demo-v20}(3)$  & 0.0355  & 0.929  & 479.8  & 0.0715  & 0.729  & 0.04946  \\ \hline 
$\mathit{demo-v20}(4)$  & 0.0634  & 3.52  & t/o  & 0.168  & 2.14  & 0.08163  \\ \hline 
$\mathit{demo-v21}(2)$  & 0.00909  & 0.0216  & 0.09432  & 0.0148  & 0.0216  & 0.584  \\ \hline 
$\mathit{demo-v21}(3)$  & 0.0283  & 0.136  & 2.672  & 0.0714  & 0.156  & 27.07  \\ \hline 
$\mathit{demo-v22}(1)$  & 0.00896  & 0.0302  & 0.104  & 0.0244  & 0.116  & 0.0816  \\ \hline 
$\mathit{demo-v22}(2)$  & 0.0107  & 0.0504  & 0.2223  & 0.0371  & 0.138  & 0.1752  \\ \hline 
$\mathit{demo-v22}(3)$  & 0.0127  & 0.0683  & 1.801  & 0.0458  & 0.341  & 0.2638  \\ \hline 
$\mathit{demo-v22}(4)$  & 0.0138  & 0.0817  & 13.89  & 0.0519  & 0.522  & 0.4677  \\ \hline 
$\mathit{demo-v22}(5)$  & 0.0153  & 0.154  & 4.138  & 0.0547  & 0.608  & 0.7734 \end{tabular}
\end{table}

\begin{table}
\caption{Computation time comparison for the Bounded Synthesis benchmarks, with modification to allow for sparser strategies. All times are given in seconds.}
\label{table:computationTimeModified}
\scriptsize\centering
\begin{tabular}{l||c|c|c|c|c} \textbf{Benchmark} & \textbf{Random} & \textbf{ILP} & \textbf{OPTSAT} & \textbf{Random} & \textbf{RepLP}  \\ & \textbf{(dumb)} & & & \textbf{(smart)} & \\ \hline
$\mathit{BasicMutex}'(3)$  & 0.0119  & 0.201  & 0.3138  & 0.0177  & 0.0528  \\ \hline 
$\mathit{BasicMutex}'(4)$  & 0.0272  & 1.83  & 1.118  & 0.107  & 0.224  \\ \hline 
$\mathit{BasicMutex}'(5)$  & 0.0728  & 20.1  & 1.535  & 0.66  & 2.06  \\ \hline 
$\mathit{ThreePhilosophers}'(3)$  & 0.0109  & 0.0628  & 0.03041  & 0.018  & 0.0426  \\ \hline 
$\mathit{ThreePhilosophers}'(4)$  & 0.0525  & t/o  & 1.155  & 0.329  & 1.09  \\ \hline 
$\mathit{ThreePhilosophers}'(5)$  & 0.197  & t/o  & 5.147  & 4.6  & 14.4  \\ \hline 
$\mathit{demo-v8}'(3)$  & 0.0143  & 0.0491  & 0.06251  & 0.0241  & 0.0456  \\ \hline 
$\mathit{demo-v8}'(4)$  & 0.0333  & 0.365  & 0.1345  & 0.161  & 0.193  \\ \hline 
$\mathit{demo-v8}'(5)$  & 0.055  & 1.6  & 0.5534  & 0.867  & 1.51  \\ \hline 
$\mathit{demo-v9}'(1)$  & 0.0218  & 12  & 0.7623  & 0.0675  & 0.466  \\ \hline 
$\mathit{demo-v9}'(2)$  & 0.477  & t/o  & 130.8  & 15.2  & 248  \\ \hline 
$\mathit{demo-v13}'(3)$  & 0.0246  & 0.0773  & 0.05977  & 0.136  & 0.0856  \\ \hline 
$\mathit{demo-v13}'(4)$  & 0.0666  & 0.745  & 0.1788  & 0.973  & 0.814  \\ \hline 
$\mathit{demo-v13}'(5)$  & 0.183  & 2.69  & 0.9863  & 2.94  & 3.32  \\ \hline 
$\mathit{demo-v15}'(1)$  & 0.0077  & 0.0742  & 0.08062  & 0.0101  & 0.0246  \\ \hline 
$\mathit{demo-v15}'(2)$  & 0.0235  & 44.1  & 5.561  & 0.0674  & 0.477  \\ \hline 
$\mathit{demo-v19}'(1)$  & 0.131  & 293  & 39.27  & 2.59  & 13.5  \\ \hline 
$\mathit{demo-v23}'(1)$  & 0.335  & t/o  & 33.45  & 14.1  & 16.9  \end{tabular}
\end{table}

\subsection{Robustness of the Approaches}

So far, we have only been concerned with the mean strategy densities (or sizes for non-positional strategies) and computation times. For practical use, it is also of importance that the fluctuations in both of these values are as low as possible. As we ran all benchmark/tool combinations 5 or 25 times, we can analyse the standard deviation for the strategy densities and times here.

In terms of strategy size, the ILP and and OPTSAT approaches have no fluctuations (as they are precise), and the strategy densities stemming from the purely random approach have a high standard deviation of up to 375 for $\mathit{demo-v19}'(1)$. However, typically, this value is between 2 and 50. The learning approach works deterministically and always returned the same result for an input safety game. For all benchmarks except for $\mathit{demo-v16}$ to $\mathit{demo-v20}$ and $\mathit{demo-v22}$, the standard deviations for the smart randomized approach are below 8, and for the repetitive LP approach, below $2.1$. For these benchmarks, the repetitive LP approach appears to be more robust, as demonstrated by Table \ref{tab:fluctuationSize}, which shows the standard deviations for $\mathit{demo-v16}$ and $\mathit{demo-v17}$ as examples. 

As far as the time is concerned, all approaches except for the smart randomized one are quite robust and have standard deviations in their computation times that are typically lower than five percent of the mean computation times, except for very small benchmarks. The smart randomized approach also has a low standard deviation in the computation times, except for the benchmarks with a high fluctuation in the strategy densities. For example, for $\mathit{demo-v20}$ through $\mathit{demo-v22}$, the standard deviation is about 20 percent of the mean computation time for the larger bounds.

\begin{table}
\caption{Standard deviations of the strategy densities/sizes for a selection of benchmarks}
\label{tab:fluctuationSize}
\scriptsize\centering
 \begin{tabular}{l||c|c|c|c|c|c} \textbf{Benchmark} & \textbf{Random (dumb)} & \textbf{ILP} & \textbf{OPTSAT} & \textbf{Random (smart)} & \textbf{RepLP} & \textbf{Learning} \\ \hline
$\mathit{demo-v16}(2)$  & 10.59  & 0  & 0  & 11.54  & 0.7483  & 0  \\ \hline 
$\mathit{demo-v16}(3)$  & 39.88  & 0  & 0  & 16.98  & 2.608  & 0  \\ \hline 
$\mathit{demo-v16}(4)$  & 47.57  & 0  & t/o  & 40.67  & 4.147  & 0  \\ \hline 
$\mathit{demo-v16}(5)$  & 94.58  & t/o  & t/o  & 48.64  & 2.638  & 0  \\ \hline 
$\mathit{demo-v17}(2)$  & 11.42  & 0  & 0  & 9.074  & 0  & 0  \\ \hline 
$\mathit{demo-v17}(3)$  & 31.5  & 0  & 0  & 16.21  & 2.245  & 0  \\ \hline 
$\mathit{demo-v17}(4)$  & 50.51  & 0  & t/o  & 38.12  & 0.4899  & 0  \\ \hline 
$\mathit{demo-v17}(5)$  & 63.23  & 0  & t/o  & 42.29  & 1.939  & 0  \end{tabular}
\end{table}

\section{Conclusion}

We performed an experimental evaluation of currently available methods to obtain sparse winning positional strategies from safety games, and compared positional strategy finding against a recent computational learning approach for non-positional strategies. 

The evaluation shows that for the explicitly represented games that stem from synthesis problems, precise methods such as applying the OPTSAT tool, or using an ILP solver is competitive in terms of computation time, although for the larger benchmarks, heuristic methods may be the only sensible way to go. For the heuristic methods, the smarter version of the randomized method is surprisingly good and comparable to the repetitive linear programming approach in terms of quality of the results. A possible reason for this good performance is that both method always find local optima. 

The learning approach to obtain non-positional strategies has shown its potential. While for most benchmarks, the strategies found by the approach were much larger than the densities of the positional ones, for others, a non-positional strategy representation that is much smaller than the density of the sparsest positional strategy was found.

For this paper, we have deliberately taken a relatively simple game model: explicit safety games. The results at hand however induce implications for more complex game types, such as symbolically represented safety or parity games: the good performance of heuristics that find local minima shows that the idea, although simple, has some potential, and gives rise to the question how this idea can be transferred to the world of symbolically represented games.

\subsection*{Acknowledgements}

The authors want to thank Daniel Neider for providing us with a copy of the implementation of the learning-based strategy extraction approach.

\bibliographystyle{eptcs}
\bibliography{bib}
\end{document}